# Research of shock waves near to the area of cumulation

G.V. Gelashvili, M.O. Mdivnishvili, I.S. Nanobashvili, S.I. Nanobashvili, G.I. Rostomashvili.
E. Andronikashvili Institute of Physics of Georgian Academy of Sciences

The purpose of the given work is detailed research of toroidal shock wave movement process to the center of symmetry in air by normal atmosphere pressure. The wave is generated by plazma which is generated by a ring discharger. The diameter of a ring is 10 cm. The capacity of the condenser battery is 1,5 mkF, the voltage of the battery was 30 KV.

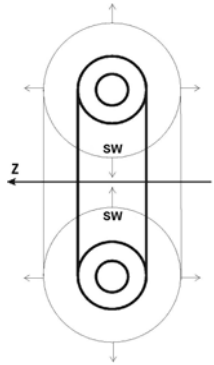

Registration of shadow pictures was carried out by means of the pulse ruby laser. Duration of a pulse is 20 ns. In Fig 1. is shown the a ring discharger and a laser beam direction.

On pictures Fig 2. stages of a shock wave convergence are shown. The moment of picture registration is specified in microseconds. It is precisely seen the front of a shock wave. It is the front of the internal side of the tore. At the moment of time 76 mks occurs a collapse of the internal side of a tore. Since this moment two variants of shock waves interaction development can occur: regular and irregular interaction. As it has been shown in earlier works (1-4), interaction begins as regular and turns in irregular at achievement of critical value of an interaction angle of the shock waves. In this case in a shadow photo the edge of the shock wave front at regular reflection should be fixed. The position of fronts is shown in Fig 3. The case of regular interaction (1) is shown by a dotted line.

Fig 1.

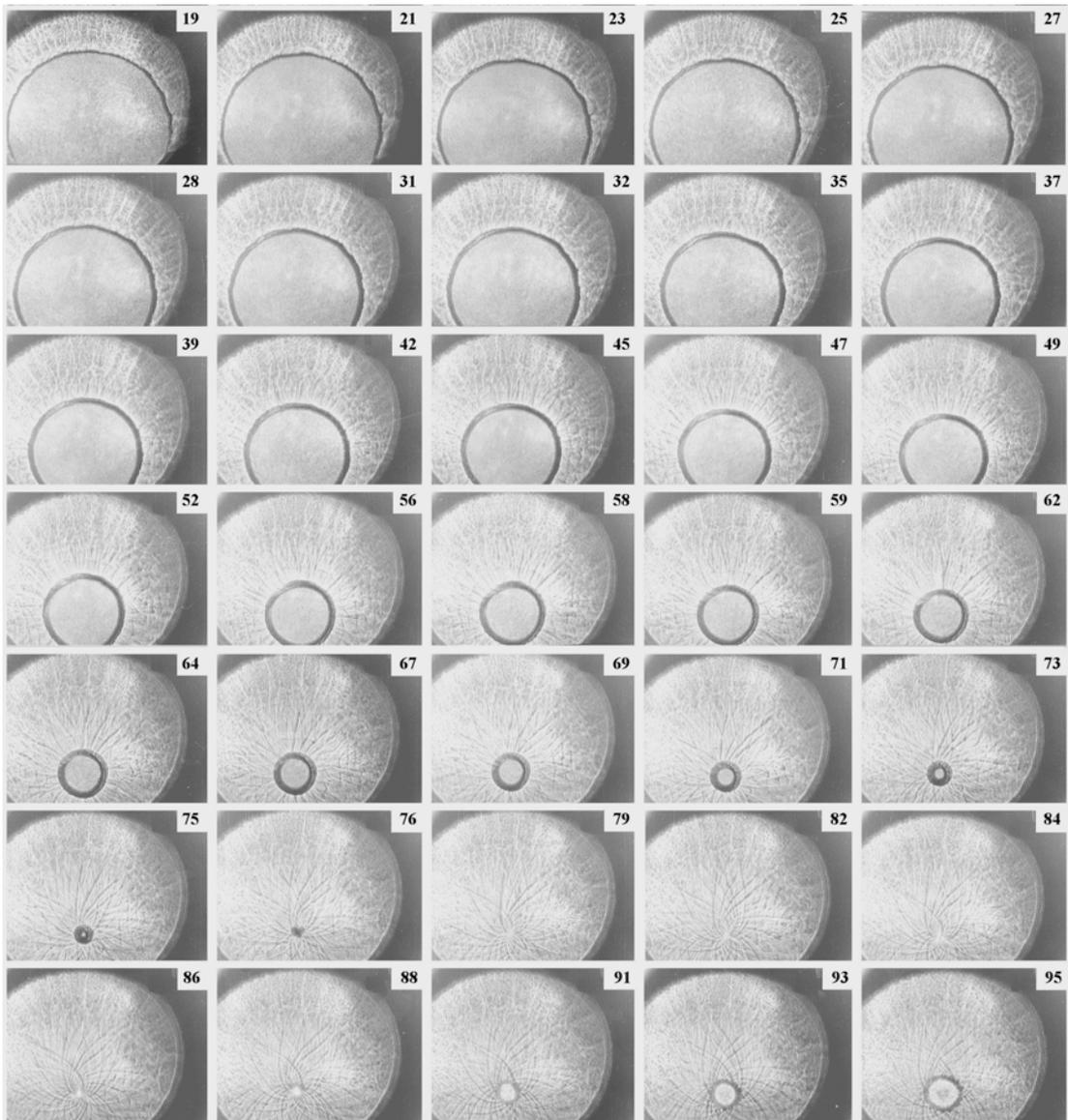

Fig 2.

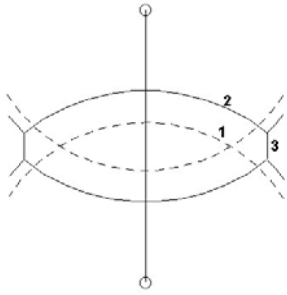

The continuous line (2) shows the front of a shock wave at irregular interaction. In this case Mach waves (3) are formed. Mach Waves move along Z axis in opposite directions. In shadow photos it should look as follows: After the moment of collapse, when the shadow of a shock wave turns to a point, the ring shadow should be appear. This shadow should correspond to the front of a shock wave at regular interaction (1). Through the certain time interval, the corner between waves 1 becomes critical. Since this moment Mach waves (3) are formed. In a shadow photo it should look as appearance of the second ring. This shadow must be formed from a triple point of interaction. The triple point is formed by two falling waves and a Mach wave.

Fig 3.

Let's consider graph Fig 4. Distances from internal edge of a ring discharger up to internal edge of a toroidal shock wave (Series 1) are shown. The data correspond to measurements submitted on Fig 2. The mark of 5cm corresponds to an axis of symmetry. For a case of regular interaction the trajectory of the chosen point at the front of shock wave should dispose on extrapolation line (Series 1). After a time interval the shadow which is formed by a triple point of interaction should appear. The Mach wave edge position (triple point) is submitted by points Series 2. For convenience of comparison the zero radius is combined with an axis of symmetry. Strictly speaking, since this moment extrapolation for Series 1 will not be correct any more. The matter is that the part of energy from the incident shock wave is redistributed in a Mach wave. In this case the speed of an initial shock wave should decrease faster, than in case of regular interaction.

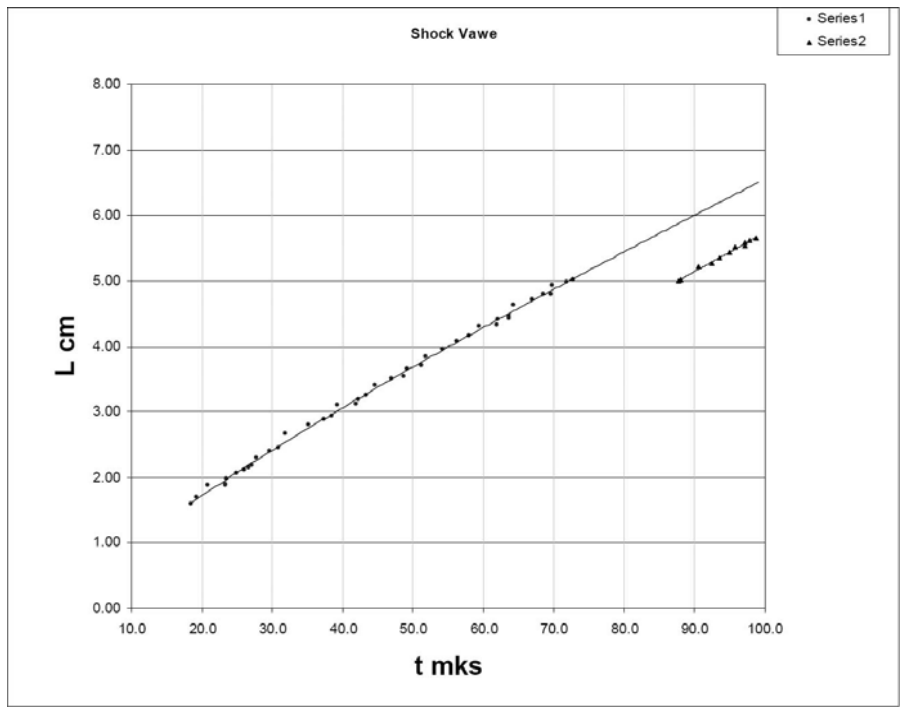

Fig 4.

Let's consider a series of pictures Fig 2. Through 10 micro seconds after the moment of the collapse, a formation of a ring shadow (the moment of time 86 mks) is observed. It cannot be an edge of the wave at regular interaction because the corresponding points on Fig 4. are located too far from extrapolation curve. Thus, on shadow pictures we clearly see a converging shock wave and formation of a Mach wave. Feature of the given pictures is that we do not see a ring shadow from a wave at regular interaction. This phenomenon is represented strange enough. It is possible to assume, that the picture "is covered" because of turbulence the rest after passage of a converging shock wave. On the other hand, the ring shadow of irregular interaction is distinctly visible on a background of turbulence. There can be two more reasons why the wave is not registered by the shadow fotography: The speed of the wave becomes less than 1 Mach that causes its transformation to a sound wave. Other reason can be in stratification of the front of a shock wave after cumulative compression. It is expressed in transformation of well-defined front of a shock wave into sequence concerning weak shock waves in the central part of cumulative interaction. In this case the method of shadow photographing can not fix edge of similar structure.

Let's consider the speed of shock wave front movement (Fig 5. ). The speed of a shock wave was calculated by means of interpolating curve constructed on experimental points of front position. Apparently from the graph, the speed of a shock wave during the initial moment of registration exceeds 2 Mach and decreases to the center, falling up to 1,7 Mach. Possessing such speed the shock wave cannot convert in a sound wave during the cumulation. Absence of its shadow after the moment of cumulation most likely is involved to occurrence of extended structure of supersonic compression waves.

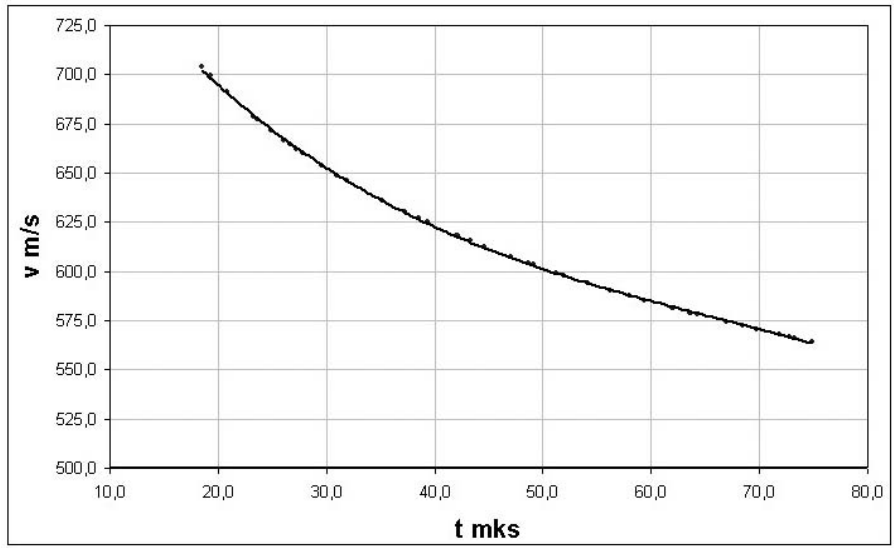

Fig 5.

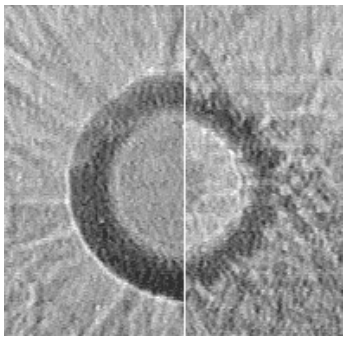

Fig 6.

Other feature is more light area in the central part of the image after a cumulation. On Fig 6. two pictures are combined. At the left is the picture up to a cumulation, on the right is the picture after formation of a Mach wave. Both pictures are made with an identical exposition. For convenience of comparison are chosen the moments of time when diameters of a ring shadows are approximately identical. It is visible, that area inside a ring in a right picture brighter than on left picture. As the image is positive, it means that the film on the right receives more light, than at the left. This phenomenon is very strange, as at the left light from the pulse ruby laser passes untroubled environment, and on the right – on the turbulent environment. It is possible to assume, that the area inside structure of shock waves differs on density. However, it is not clear, why the light beam brightness increases.

To determine dynamics of plasma during the discharge, photo filming has been carried out by means of a high-speed camera. Apparently from pictures Fig 7., the maximal luminescence of plasma is observed in a time interval 2.2 – 4.3 mks. By the moment of time 31.6 mks. plasma practically dies out. Apparently from pictures, plasma formation does not reach the center of a ring. It is necessary to note one more circumstance. Comparing shadow photos Fig 2. and photos of a luminescence of plasma Fig 7. it is necessary to pay attention at the moment of time equal 19.0 mks. As we can see on a picture of a luminescence, plasma is near to a surface of a ring and its distance from the center makes not less than 40 mm, the distance from the center of a ring up to a shock wave makes 34 mm. Thus, by this moment of time the shock wave has already come off plasma. It is necessary to note one more important detail. The shock wave is formed at edge of plasma formation. Thus, the edge of plasma cannot be closer than

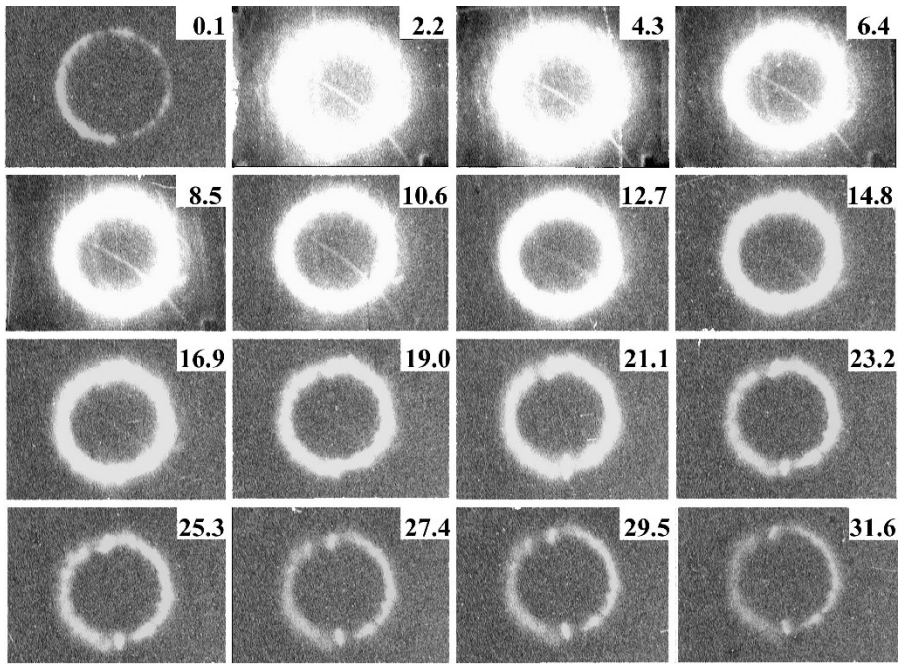

Fig 7.

34 millimeters from the center of the ring. From here follows, that seen approach of the luminescence of plasma time of 2.2 mks is optical effect. In reality plasma formation does not come to the center of a ring closer than 34 millimeters. Moreover, by the time of 32 mks plasma practically dies of.

**Summary:**

The specified measurements of a converging shock wave speed are carried out. It is shown, that the front of a shock wave is not fixed after interaction on an axis of symmetry.

The effect of enlightenment in the central area after formation of a Mach wave is found out.

It is shown, that plasma of the toroidal discharge does not reach the center of a ring under the given conditions of experiment, and dies out long before the moment of a shock wave cumulation in the center. Thus plasma cannot accept participation in effects developing after cumulation.